\documentclass[aps,prl,numerical,linenumbers,superscriptaddress,showpacs,floatfix,reprint]{revtex4-1}
\usepackage[dvipdfm]{graphicx}% Include figure files
\usepackage{amsmath,amssymb,amsfonts}
\usepackage[colorlinks=true,linkcolor=blue,plainpages=false,hypertex]{hyperref}

\def\v2{\mbox{$v_2$}}

\begin{document}

%\pagewiselinenumbers

%%%%%%%%%%%%%%%%%%%% Title and Authors %%%%%%%%%%%%%%%%%%%%%%%%%%%%%%%%%%%%%%
%
\title{ Glauber-based evaluations of the odd moments of the initial eccentricity \\
relative to the even order participant planes 
}
%
%****************************  New Feature ******************************
%%%%%%%%%%%%%%%%%%%%%%%%%%%%%%%%%%%%%%%%%%%%%%%%%%%% PRL length check lines

%\author{ Authors }
\author{ Roy~A.~Lacey}
\email[E-mail: ]{Roy.Lacey@Stonybrook.edu}
\affiliation{Department of Chemistry, 
Stony Brook University, \\
Stony Brook, NY, 11794-3400, USA}
%\affiliation{Physics Department, Bookhaven National Laboratory, \\
%Upton, New York 11973-5000, USA}
\author{Rui Wei} 
\affiliation{Department of Chemistry, 
Stony Brook University, \\
Stony Brook, NY, 11794-3400, USA}
\author{ N.~N.~Ajitanand} 
\affiliation{Department of Chemistry, 
Stony Brook University, \\
Stony Brook, NY, 11794-3400, USA}
\author{ J.~M.~Alexander}
\affiliation{Department of Chemistry, 
Stony Brook University, \\
Stony Brook, NY, 11794-3400, USA}
%\author{ X.~Gong}
%\affiliation{Department of Chemistry, 
%Stony Brook University, \\
%Stony Brook, NY, 11794-3400, USA}
\author{ J.~Jia}$^2$
\affiliation{Department of Chemistry, 
Stony Brook University, \\
Stony Brook, NY, 11794-3400, USA}
\affiliation{Physics Department, Bookhaven National Laboratory, \\
Upton, New York 11973-5000, USA}
\author{A.~Taranenko}
\affiliation{Department of Chemistry, 
Stony Brook University, \\
Stony Brook, NY, 11794-3400, USA} 

%\author{\\ R. Pak}
%%
%\affiliation{Physics Department, Bookhaven National Laboratory, \\
%Upton, New York 11973-5000, USA}

%\author{Horst St\"ocker}
%%
%\affiliation{Institut f\"ur Theoretische Physik, Johann Wolfgang Goethe-Universit\"at \\
%             D–60438 Frankfurt am Main, Germany} 

\date{\today}

%%%%%%%%%%%%%%%%%%%%%%%%%%%%%% Abstract %%%%%%%%%%%%%%%%%%%%%%%%%%%%%%%%%%%%%%

\begin{abstract}

	Monte Carlo simulations are used to compute the centrality dependence of 
the odd moments of the initial eccentricity $\varepsilon_{n+1}$, relative 
to the even order (n) participant planes $\Psi_n$ in Au+Au collisions. The results 
obtained for two models of the eccentricity -- the Glauber and the factorized 
Kharzeev-Levin-Nardi (fKLN) models -- indicate magnitudes which are essentially 
zero. They suggest that a possible correlation between the orientations of the 
the odd and even participant planes ($\Psi_{n+1}$ and $\Psi_n$ respectively), 
do not have a significant influence on the calculated eccentricities. An experimental
verification test for correlations between the orientations of the 
the odd and even participant planes is also proposed.

%reaction zone produced in 
%collisions between two heavy nuclei, 
 
\end{abstract}


\maketitle

%%%%%%%%%%%%%%%%%%%%%%%%%%%%% Introduction %%%%%%%%%%%%%%%%%%%%%%%%%%%%%%%

	The magnitude and fluctuations of the initial eccentricity of the collision zone,  
has proven to be an essential ingredient in ongoing efforts to extract the 
transport properties of the quark gluon plasma (QGP)
\cite{Gyulassy:2000er,Wang:2001ifa,Teaney:2003kp,Lacey:2006pn,Luzum:2008cw,
Song:2008hj,Majumder:2007hx,Bass:2008rv,Dusling:2007gi,Chaudhuri:2009hj,Bozek:2009mz,
Denicol:2010tr,Holopainen:2010gz,Molnar:2001ux,Xu:2007jv,Greco:2008fs,
Drescher:2007cd,Lacey:2009xx,Masui:2009pw,Lacey:2010fe,Xu:2010cq,Lacey:2009kg,
Lacey:2009kg,Lacey:2009ps,Noronha:2009vz,Shen:2010uy,Jia:2010ee,
Adare:2010sp,Gossiaux:2010yx,Younus:2010sx}. 
Experimental measurements of this eccentricity have not been possible to date.
The necessary theoretical estimates can be obtained from Glauber-based 
models \cite{Miller:2007ri,Drescher:2007ax} via the two-dimensional profile $S$ 
of the density of sources in the transverse plane $\rho_s(\mathbf{r_{\perp}})$ of 
the overlap collision geometry specified by the impact parameter $b$, or the number 
of participants $N_{\text{part}}$ \cite{Miller:2007ri,Alver:2006wh,
Hama:2007dq,Broniowski:2007ft,Hirano:2009ah,Lacey:2009xx,Gombeaud:2009ye,
Lacey:2010yg,Schenke:2010rr,Staig:2010pn,Qin:2010pf}:
  \begin{eqnarray}  % Definition of event plane 
    S_{nx} & \equiv & S_n \cos{(n\Psi_n)} = 
    \int d\mathbf{r_{\perp}} \rho_s(\mathbf{r_{\perp}}) \omega(\mathbf{r_{\perp}}) \cos(n\phi), \label{eq:S_x} \\
    S_{ny} & \equiv & S_n \sin{(n\Psi_n)} = 
    \int d\mathbf{r_{\perp}} \rho_s(\mathbf{r_{\perp}}) \omega(\mathbf{r_{\perp}}) \sin(n\phi), \label{eq:S_y} \\
%    S_x & \equiv & S_n \cos{(n\Psi^*_n)} = \int r dr \omega_i \cos{(n\phi_i)}, \label{eq:S_x} \\
%    S_y & \equiv & S_n \sin{(n\Psi^*_n)} = \int r dr  \omega_i \sin{(n\phi_i)}, \label{eq:S_y} \\
    \Psi_n & = & \frac{1}{n} \tan^{-1}\left(\frac{S_{ny}}{S_{nx}}\right), \label{eq:S_n-plane}
  \end{eqnarray}
where $\phi$ is the azimuthal angle of each source, the 
weight $\omega(\mathbf{r_{\perp}}) = \mathbf{r_{\perp}}^2$, 
$\Psi_n$ is the azimuth of the rotation angle for the minor axis of 
the $n$-th harmonic of the shape profile, and 
\begin{eqnarray}
\varepsilon_n = \left\langle \cos n(\phi - \Psi_n) \right\rangle, \nonumber \\
\varepsilon^*_n = \left\langle \cos n(\phi - \Psi_m) \right\rangle, \, \, n \ne m
\label{e2e4}
\end{eqnarray}
are the $n$-th order moments of the eccentricity obtained relative to 
$\Psi_n$ and $\Psi_m$ respectively \cite{Broniowski:2007ft,Lacey:2010yg,Lacey:2010hw}; 
the brackets denote averaging over sources, as well as events belonging to a particular 
centrality or impact parameter range. Note that $\Psi_n$ is restricted to the range 
$0.0 - \pi/n$ radians due to its $n$ fold symmetry (separated by $2\pi/n$).

For such estimates, the geometric fluctuations associated with the positions of the nucleons are a primary source of the initial eccentricity fluctuations. That is, for a given centrality, the fluctuating positions of the participant nucleons lead to event-by-event fluctuations of the so-called participant planes ($\Psi_n$) about the reaction plane, defined by the beam direction and the impact parameter. 
An obvious consequence of these fluctuations is that the participant eccentricities $\varepsilon_n$ are larger than the so-called standard eccentricities, evaluated  relative to the reaction plane. The difference between the standard and participant eccentricities is of course centrality dependent and can be relatively sizable for central and peripheral 
collisions.
%eccentricity ($\varepsilon_2^*$) referenced to this plane.

	Trivial auto-generated correlations are also inherent in Glauber-based models. Such correlations stem from the fact that a single nucleon from nucleus $A$ ``wounds'' several nucleons from nucleus $B$, when the two collide. Thus, a certain degree of clustering or correlations between the locations of ``wounded'' nucleons is expected to be generated in collisions. These correlations are tantamount to a decrease in the effective number of sources in the collision zone, so they are expected to lead to a small [centrality dependent] increase in the magnitudes for $\varepsilon_n$. The scaled fluctuations $\Delta \varepsilon_n/\varepsilon_n$  
show a more complicated centrality dependence but are insensitive to 
the correlations in the most central events \cite{Broniowski:2007ft}. 
Another potential influence of the auto-generated correlations is that they 
could induce a correlation between the participant planes for the 
even ($\Psi_n$) and odd ($\Psi_{n+1}$) eccentricity moments 
(especially in peripheral events) and hence, influence their relative magnitudes. 
Thus, an important question is the degree to which such correlations influence 
the extracted values for $\varepsilon_n$ (for odd and even $n$) \cite{Nagle:2010zk}?

	A simple approach to evaluate this influence, is to compute the odd 
eccentricity moments $\varepsilon_{3,5,...}$ with 
respect to the even order participant planes $\Psi_{2,4,...}$. 
%\cite{restrict_plane}. 
Here, the essential point 
is that, a significant correlation between $\Psi_2$ and $\Psi_{3,5,}$ [for example] should lead to sizable values for $\varepsilon_{3,5}$. On the other hand, if the computed values for $\varepsilon_{3,5,...} \approx 0$ then, for all intent and purposes, $\Psi_n$ and $\Psi_{n+1}$ can be considered to be uncorrelated, as has been claimed in several recent papers (see for example 
Refs. \cite{Alver:2010gr,Staig:2010pn,Qin:2010pf,Lacey:2010hw}). 
Note that for symmetric collisions with smooth eccentricity profiles, the distributions are symmetric under the transformation $\Psi_2 \rightarrow \Psi_2 + \pi$ so all odd harmonics are identically zero. However, the ``lumpy'' collision zones considered here do not have any particular symmetry and the odd harmonics are not required to be zero from event to event. 

%
%%%%%%%%%%%%%%%%%%%%%%%%%%%%%%%%  Figure 1  %%%%%%%%%%%%%%%%%%%%%%%%%%%%%%%%
\begin{figure}[tb]
%\begin{tabular}{cc}
\includegraphics[width=1.0\linewidth]{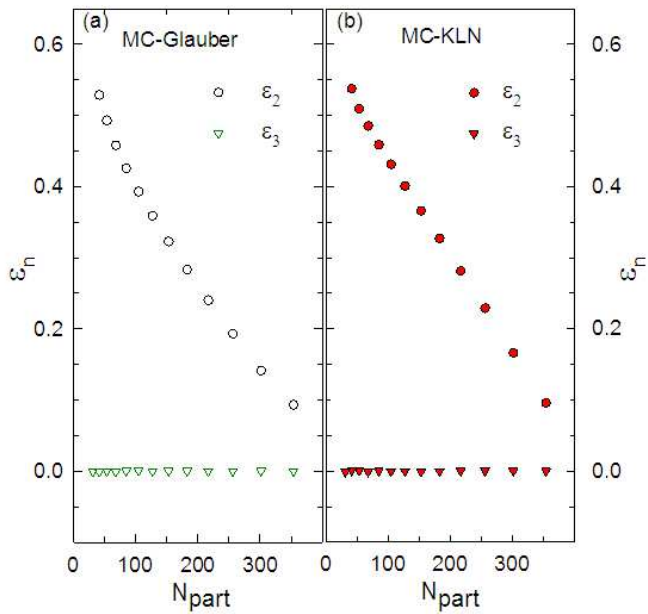}
%\includegraphics[width=0.5\linewidth]{v2rpRes.eps} &
%\includegraphics[width=0.5\linewidth]{v4rpRes.eps}
%\end{tabular}
%
\caption{ Comparison of $\varepsilon_{2}$ and $\varepsilon_{3}$ vs. $N_{\rm part}$, 
for Au+Au collisions. The odd eccentricity moments are evaluated 
with respect to the $\Psi_2$ participant plane.
The open and filled symbols show the results from MC-Glauber and MC-KLN as indicated.
}
\label{Fig1}
\end{figure}

	Monte Carlo calculations were performed following the implementation 
scheme outlined in Refs. \cite{Lacey:2010yg,Lacey:2010hw} for the 
Glauber (MC-Glauber) \cite{Alver:2006wh,Miller:2007ri} and the factorized 
Kharzeev-Levin-Nardi (MC-KLN) models \cite{Lappi:2006xc,Drescher:2007ax}. 
A subset of the results from these calculations is shown in Fig. \ref{Fig1};      
it shows calculated values for $\varepsilon_{2}$ vs. 
$N_{\text{part}}$ and $\varepsilon_{3}$ vs. $N_{\text{part}}$ for Au+Au collisions. 
The reader is reminded here that both $\varepsilon_2$ and 
$\varepsilon_{3}$ are computed relative to the $\Psi_2$ participant plane.
For $\varepsilon_{3}$ the five-particle correlator 
$\left< cos\left( 3\phi_1 + 3\phi_2 - 2\phi_3 - 2\phi_4 -2\phi_5 \right) \right>/\varepsilon_2^3 = \varepsilon^2_{3/\Psi_2}$
can be used.

	The open and filled triangles in Fig. \ref{Fig1} indicate that the values for the odd moments, 
obtained for both MC-Glauber and MC-KLN, remain flat as a function 
of collision centrality and are essentially equal to zero. 
Here, it is noteworthy that the event-by-event fluctuations of $\Psi_{3}$ about $\Psi_2$ 
lead to a broad distribution of $\varepsilon_{3}$ values which range from 
negative to positive values. Thus, when averaged over events, they give
magnitudes $\approx 0$. Note as well that these magnitudes are 
minuscule when compared to the participant eccentricities $\varepsilon_{3}$, 
calculated with respect to $\Psi_3$ \cite{Lacey:2010hw}
and $\varepsilon^*_{4}$ calculated with respect to $\Psi_2$ \cite{Lacey:2010yg}. 
These results show that our eccentricity evaluations suffer little, if any, 
influence from possible correlations between the odd and even 
participant planes.

	The magnitudes and trends for $\varepsilon_n$ 
are expected to influence the magnitude and trends for anisotropic 
flow \cite{Lacey:2010yg,Alver:2010gr,Alver:2010dn,Holopainen:2010gz,
Petersen:2010cw,Staig:2010pn,Qin:2010pf,Schenke:2010rr,Lacey:2010hw},
characterized by the Fourier coefficients $v_{n}$. Consequently, our 
approach can be used to perform actual experimental tests for correlations 
between the odd and even participant planes. That is, 
an experimental estimate can be obtained by measuring the 
Fourier coefficients $v_{n+1}$ ($v_{n}$) with respect to the 
$\Psi_n$ ($\Psi_{n+1}$) participant planes. Similarly, a 
direct experimental measurement of the correlation between the 
odd ($\Psi_{n+1}$) and even ($\Psi_{n}$) event planes can be made. 
%over the range $0.0-2\pi/(2(n+1))$ radians.

%%%%%%%%%%%%%%%%%%%%%%%%%%%%%  Conclusions  %%%%%%%%%%%%%%%%%%%%%%%%%%%%%%

In summary, we have presented results for the odd initial eccentricity moments 
$\varepsilon_{n+1}$, determined relative to the even order $\Psi_n$ planes for 
Au+Au collisions, for two primary models. The calculated values 
of $\varepsilon_{n+1}$ are found to be essentially zero, indicating the 
absence of any significant influence from a possible correlation between 
the odd and even order participant planes, inherent in Glauber-based models. 
This finding reaffirms earlier conclusions that, for 
eccentricity evaluations, the odd and even order participant 
planes ($\Psi_n$ and $\Psi_{n+1}$) can be taken to be uncorrelated.
It remains to be seen whether these findings are supported by actual 
experimental measurements.

%%%%%%%%%%%%%%%%%%%%%%%%%%  Acknowledgements  %%%%%%%%%%%%%%%%%%%%%%%%%%
%\section*{Acknowledgments}
{\bf Acknowledgments}
%%%%%%%%%%%%%%%%%%%%%%%%%%%%%%%%%%%%%%%%%%%%%%%%%%%%%%%%%%%%%%%%%%%%%%%%%%%%%%%%%%%%%%%%%%%%%%%%%%%%%%%%%%%%%%%%%%%%%%%%
%We thank Paul Mantica (MSU/NSCL) for crucial insights on nuclear deformation.
This research is supported by the US DOE under contract DE-FG02-87ER40331.A008 
 
%%%%%%%%%%%%%%%%%%%%%%%%%%%%%%%%%%%%%%%%%%%%%%%%%%%%%%%%%%%%%%%%%%%%%%%%%%%%%%%%%%%%%%%%%%%%%%%%%%%%%%%%%%%%%%%%%%%%%%%%%

%%%%%%%%%%%%%%%%%%%%%%%%%%%%%  References  %%%%%%%%%%%%%%%%%%%%%%%%%%%%%%

%
\bibliography{ecc_fluc_x0} 
\end{document}